# Decentralized On-Ramp Merging Control of Connected and Automated Vehicles in the Mixed Traffic Using Control Barrier Functions

Haoji Liu, *Student Member, IEEE*, Weichao Zhuang, *Member, IEEE,*
Guodong Yin*, *Senior Member, IEEE,* Rongcan Li, Chang Liu and Shanxing Zhou

*Abstract*— The cooperative control of the connected and automated vehicle (CAV) is recognized as an effective approach to alleviate traffic congestion and improve traffic safety, especially for on-ramp bottlenecks. However, in the mixed traffic, the uncertainty of human-driven vehicles (HDVs) makes the on-ramp merging control for CAVs more challenging. This paper proposes a decentralized optimal control method to address the merging control problem of CAVs at highway on-ramps in the mixed traffic. We first formulate the optimal merging control problem, which includes the constraints of safety and vehicle dynamics, with the objectives of minimizing travel time and energy consumption. Then, a control framework, combining control barrier functions (CBFs) and control Lyapunov functions (CLFs) is proposed. CBFs render the system subject to safety-critical constraints, while CLFs stabilize the system to the objectives. In addition, to enable effective control of CAVs in the mixed traffic, a recursive merging control framework is proposed, where HDVs are regarded as disturbances, and CAVs collect surrounding vehicles' states repeatedly and update their trajectories recursively to satisfy strict merging requirements. Finally, the merging problem is reformulated as a quadratic programming problem, which allows for real-time application. Simulation results show that the proposed on-ramp merging control method is robust in resisting disturbance from the HDV with traffic efficiency and energy economy improvement.

## I. Introduction

The on-ramp merging is one of the main triggers of traffic congestion on highways. Such bottlenecks can cause speed breakdown, capacity drop, and further leading to excessive fuel consumption [1]. In addition, on-ramp merging is a common but stressful task for drivers due to the involved risk and the close interaction with other drivers. Traffic perturbations, inefficient traffic flow, and even crashes may happen in the on-ramp merging maneuver due to different abilities, characteristics, and driving styles of interacting vehicles and drivers. With the emerging of Connected and Automated Vehicles (CAVs), regulating individual vehicle's trajectory becomes possible which provides a new approach to improve traffic efficiency and safety simultaneously at highway on-ramps.

The cooperative merging control of CAVs can be broadly separated into centralized and decentralized control [2]. For the centralized control, vehicles in the merging zone are controlled by a central controller, which calculates the trajectory of each vehicle in the control zone by minimizing the global cost. Rios-Torres *et al*. [3] transformed the centralized merging control problem into an unconstrained optimal control problem, and a closed-form analytical solution was given using Hamiltonian analysis. Jing *et al*. [4] proposed a centralized optimization framework based on cooperative game theory and the Pareto efficient solution which minimized the global cost was obtained to optimize merging sequence and trajectory. In addition, Zhao *et al*. [5] designed a hierarchical merging system, where the upper layer aims to optimize the macroscopic traffic flow, and the lower layer calculates the control inputs of vehicles based on planning results from the upper layer.

Although the centralized control strategy can obtain a globally optimal solution, it may suffer great computational burdens, especially for heavy traffic. The decentralized control strategy may be an effective approach, in which each vehicle plans its own trajectory by using shared traffic information. Hayashi *et al*. [6] proposed a decentralized model predictive control strategy, and used potential function to avoid vehicle collision in the merging zone. Fukuyama [7] formulated the ramp merging problem as a dynamic game problem between vehicles, and an efficient calculation method, zero-suppressed binary decision diagram, was used to derive the optimal merging trajectory.

Most existing on-ramp merging control studies focus on the scenario where all vehicles are CAVs. However, the vehicle without connectivity and automation will not be replaced overnight. Thus, it is important to investigate the merging control strategy of CAVs in the mixed traffic where CAVs and human-driven vehicles (HDVs) coexist [8]. In the mixed traffic, HDVs are not controlled by advanced controller and their information cannot be accurately predicted in advance, which can be regarded as a disturbance for CAVs. As each CAV can plan its own movement independently using the decentralized control strategy, which has a good potential to resist HDVs' interference, the ramp merging problem in the mixed traffic is often constructed as a decentralized control problem. Liao *et al*. [9] introduced an agent-based game method to optimize the merging sequence and trajectory of CAVs in the mixed traffic. Karimi *et al*. [10] focused on the merging behavior of three vehicles and obtained control

Research supported by Key R&D Program of Jiangsu Province under grant BE2019004, Achievements Transformation Project of Jiangsu Province under Grant BA2018023, National Natural Science Foundation of China under Grants 51975118 and 52025121. *Corresponding author: Guodong Yin.*

The authors are with School of Mechanical Engineering, Southeast University, Nanjing, Jiangsu, 211189 China (e-mail: hjl@seu.edu.cn; wezhuang@seu.edu.cn; ygd@seu.edu.cn; lirongcan@seu.edu.cn; me_chang liu@seu.edu.cn; shiningstar@seu.edu.cn).

decisions based on model predictive control framework. However, the motion behavior of HDVs is simplified using certain models, which cannot reflect the uncertain maneuver of HDVs.

In order to achieve the optimal on-ramp merging control of CAVs in the mixed traffic, this paper introduces a recursive optimal on-ramp merging control method, which combines the control barrier functions (CBFs) and control Lyapunov functions (CLFs). The combination of CBFs and CLFs realizes the unification of safety and stability, especially suitable for the safety-critical controller with optimization objectives [11, 12]. Recently, Xiao et al. [13] also used this method to achieve cooperative merging control of CAVs in full CAV traffic.

The main contributions of this paper are threefold. First, HDVs in the mixed traffic are regarded as random disturbances to CAVs, and the decentralized control strategy is adopted to realize the dynamic trajectory planning and control while dealing with the disturbances. Second, a CBF-CLF based optimal control method is introduced, and the control problem is transformed into a discrete quadratic programming problem, which can realize the optimal control of the CAV's trajectory under complex constraints in the mixed traffic. Third, the recursive optimal control framework for CAVs is designed to make CAVs have the ability to resist uncertain disturbances of HDVs in real time.

The remainder of this paper is organized as follows. In Section II, we formulate the problem of on-ramp merging in the mixed traffic, and give the control objectives and constraints for CAVs. In Section III, a recursive optimal control framework is proposed, and a CBF-CLF based merging controller is designed. Simulation results and analysis are presented in Section IV, and conclusions are stated in Section V.

## II. PROBLEM FORMULATION

### A. Merging Scenario Description

We consider a common merging scenario with a one-lane main road and a one-lane on-ramp connected with an acceleration lane as shown in Figure 1. The merging point is specified close to the end of the acceleration lane, at which all the vehicles are enforced to perform the final lateral motion to merge into the mainline. In this paper, we only consider the longitudinal motion of merging vehicles, and assume the lateral motion of merging vehicles has no influence on it. Therefore, such two movements can be treated separately. Further, within the communication coverage of the local coordinator, the coordination zone is defined and has the length of $L$.

The local coordinator plays the role of identifying vehicles, specifying the merging sequence, and broadcasting information to CAVs. At each time instant, the total number of vehicles entering the coordination zone is $N(t)$. Instead of designing the optimal merging sequence which is beyond the scope of this paper, we impose a strict First-in-first-out (FIFO) rule, i.e., each vehicle must reach the merging point in the same order it enters the coordination zone. Once a vehicle enters the coordination zone, it will be assigned a unique identity $i = N(t) + 1$ by the local coordinator. The smaller the value of the vehicle identity $i$, the earlier it reaches the merging point. In the case that a vehicle passes through the merging point, the identity of this vehicle will be eliminated, and the identities of the vehicles in the coordination zone will be reduced by 1 accordingly.

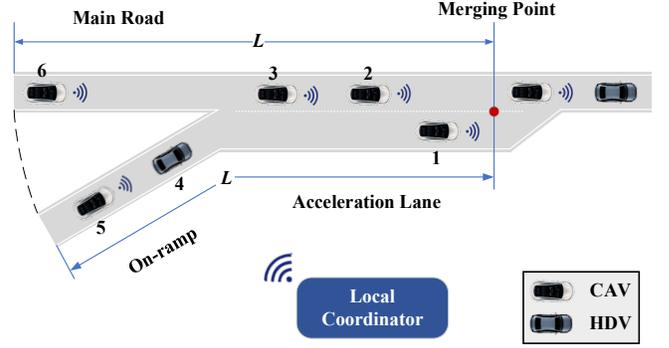

Figure 1. The mixed traffic on-ramp merging scenario.

Both the CAV and HDV will be assigned identities after entering the coordination zone (the information of HDV entering the zone can be captured by roadside units and broadcast to CAVs by the local coordinator). For vehicle $i$ ($i = 1,2,…,N(t)$), the identity of its preceding vehicle in the same lane is denoted as $ip$. Then there are two cases:

1) $ip = i - 1$ when vehicle i-1 and vehicle $i$ are in the same road, just like the relationship between vehicle 3 and vehicle 2 in Fig. 1.

2) $ip < i - 1$ if the vehicle $i - 1$ passing through the merging point immediately before vehicle $i$ is in the different road. For example, the corresponding $ip$ of vehicle 4 is 1 while $i - 1 = 3$ in Fig. 1.

In the mixed traffic, both CAV and HDV exist. The CAV can communicate with local coordinator and other CAVs through V2I and V2V communication, and obtain other CAVs' states in real time while broadcasting its own state. However, the HDV does not have the ability of communication, and its motion state can only be collected through the on-board sensors of nearby vehicles. Therefore, when a HDV is in front of a CAV, the CAV needs to use on-board sensors (such as radars) to repeatedly collect the information of the preceding HDV and adjust the distance, so as to achieve the requirements of safe merging and safe car-following.

Using the decentralized control strategy, each CAV plans its own trajectory and achieves accurate control according to the coordination information of local coordinator and the perceived surrounding traffic environment information.

### B. Vehicle Dynamics

In this paper, we consider all CAVs are homogenous electric vehicles. Vehicle parameters are listed in Table I. Only the longitudinal trajectory is considered. The longitudinal vehicle dynamics of CAV $i$ is

$$\delta M a_i(t) = F_{t,i}(t) - F_{r,i}(t) \tag{1a}$$

$$F_{r,i}(t) = Mgf_r\sin\alpha + Mg\cos\alpha + 0.5\rho C_d A v_i^2(t) \tag{1b}$$

where $a_i(t)$ and $v_i(t)$ is the acceleration and speed, $F_{t,i}(t)$ is traction force, which is the input from the motor to wheels, or longitudinal vehicle dynamics. Select $F_{t,i}(t)$ as system control input and denote it as $u_i(t)$, select position $d_i(t)$ and speed $v_i(t)$ constitutes the state vector $\boldsymbol{x}_i(t) \coloneqq (d_i(t), v_i(t))$, the nonlinear longitudinal vehicle control system is expressed as

$$\underbrace{\begin{bmatrix} \dot{d}_i(t) \\ \dot{v}_i(t) \end{bmatrix}}_{\dot{x}_i(t)} = \underbrace{\begin{bmatrix} v_i(t) \\ -\frac{1}{\delta M} F_{r,i}(t) \end{bmatrix}}_{f(\boldsymbol{x}_i(t))} + \underbrace{\begin{bmatrix} 0 \\ \frac{1}{\delta M} \end{bmatrix}}_{g(\boldsymbol{x}_i(t))} u_i(t) \quad (2)$$

TABLE I. VEHICLE PARAMETERS

| Parameters | Symbol | Value |
|---|---|---|
| Mass | $M$ | 1997 kg |
| Rotational inertia coefficient | $\delta$ | 1.05 |
| Gravity constant | $g$ | 9.81 m/s² |
| Rolling resistance coefficient | $f_r$ | 0.012 |
| Road slope | $\alpha$ | 0 |
| Air density | $\rho$ | 1.2 kg/m³ |
| Aerodynamic drag coefficient | $C_d$ | 0.22 |
| Windward area | $A$ | 2.4 m² |
| Wheel radius | $R$ | 0.34 m |
| Transmission ratio | $\mathcal{T}$ | 9.7 |
| Tunable motor parameter | $c$ | 0.8730 |

The motor power $P_{m,i}(t)$ is approximated by a widely used function of motor torque $T_{m,i}(t)$ and rotation speed $n_{m,i}(t)$ [14,15]

$$P_{m,i}(t) = n_{m,i}(t) T_{m,i}(t) + c T_{m,i}^2(t) \quad (3)$$

Neglecting motor-to-wheel transmission efficiency yields

$$n_{m,i}(t) = v_i(t) \mathcal{T}/R \quad (4a)$$
$$T_{m,i}(t) = u_i(t) R/\mathcal{T} \quad (4b)$$

Substituting (4) into (3) derives

$$\begin{aligned} P_{m,i}(t) &= \frac{v_i(t)\mathcal{T}}{R}\frac{u_i(t)R}{\mathcal{T}} + c\left(\frac{u_i(t)R}{\mathcal{T}}\right)^2 \\ &= v_i(t) u_i(t) + c' u_i^2(t) \end{aligned} \quad (5)$$

where $c$ is a tunable motor parameter, $c' = cR^2/\mathcal{T}^2$. The corresponding required battery power $P_{b,i}(t)$ is simplified ignoring auxiliary power and energy loss [14], i.e., $P_{b,i} = P_{m,i}$.

*C. Optimal Merging Control Problem Formulation*

For each CAV, the control objective is minimizing travel time and battery energy consumption during on-ramp merging control process.

*Problem I. Optimal control problem for CAV i*

$$\begin{aligned} \min J_i &= \min \int_{t_i^0}^{t_i^m} (P_{b,i} + w_i) dt \\ &= \min \int_{t_i^0}^{t_i^m} (v_i u_i + c' u_i^2 + w_i) dt \end{aligned} \quad (6a)$$

Subject to

$$a_{min} \leq a_i(t) \leq a_{max}, \forall t \in [t_i^0, t_i^m] \quad (6b)$$
$$v_{min} \leq v_i(t) \leq v_{max}, \forall t \in [t_i^0, t_i^m] \quad (6c)$$
$$z_{i,ip} = d_{ip}(t) - d_i(t) \geq l + \varphi v_i(t), \forall t \in [t_i^0, t_i^m] \quad (6d)$$
$$z_{i,i-1} = d_{i-1}(t_i^m) - d_i(t_i^m) \geq l + \varphi v_i(t_i^m) \quad (6e)$$

where $t_i^0$ is the time when vehicle $i$ enters the coordination zone, and $t_i^m$ is the time when vehicle $i$ arrives at the merging point, $w_i$ is a constant penalty for travel time. Constraint (6b) reflects driving comfort requirements [16], which gives the maximum acceleration $a_{max} > 0$ and deceleration $a_{min} < 0$, which. (6c) represents speed limitation of roads, where $v_{min} \geq 0$ is the minimum speed and $v_{max} \geq 0$ is the maximum speed. (6d) is the safe car-following constraint, i.e., CAV $i$ must have enough safety distance from its preceding vehicle $ip$ at any time $\forall t \in [t_i^0, t_i^m]$, where $l$ is a constant denoting the minimum safe distance, and $\varphi$ is the time headway, which is generally taken as 1.8. (6e) represents the safe merging constraint for CAV $i$ to keep a safe distance from vehicle $i - 1$ at the merging point. Note that, if CAV $i$ is on the on-ramp, it should always keep safe merging constraint before merging. However, for CAV $i$ on the main road, it is constrained by (6e) only when $ip < i - 1$, i.e., there is an on-ramp vehicle $i - 1$ wants to merge into the main road in front of CAV $i$.

In the mixed traffic, CAVs should always adjust their trajectories to reduce the impact of HDVs. Thus, in this paper, we introduce a recursive optimal control framework. Details are illustrated in Section III.

### III. CBF-CLF BASED CONTROLLER DESIGN

*A. Recursive Optimal Control Framework*

In the mixed traffic, the HDV cannot be accurately controlled, and its trajectory cannot be predicted in advance. Therefore, there is no guarantee that a certain trajectory of the CAV can always be feasible. To address Problem I, we propose a recursive optimal control framework for CAVs, as shown in Figure 2. In this framework, the CAV repeatedly collects the HDV's information and replans its trajectory, so as to satisfy safe constraints all the time. Specifically, the CAV needs to recursively update the speed, position and acceleration of the HDV, and use the real-time information to solve Problem I repeatedly.

Denote the recursive period as $\Delta t$. Each recursive interval for CAV $i$ is expressed as $[t_i^0 + k\Delta t, t_i^0 + (k+1)\Delta t]$, $k = 0,1,2 \ldots$. CAV $i$ plans its trajectory at the beginning of each time interval, i.e., at $t_i^0 + k\Delta t$. Here an assumption is given that the HDV keeps its speed at a constant value $v_i(t_i^0 + k\Delta t)$ in each recursive interval. Further, the control input $u_i(t_i^0 + k\Delta t)$ is calculated and applied in the recursive interval. Such a process is repeated until CAV $i$ reaches the merging point. Because the whole process is executed recursively, there is no guarantee that CAV $i$ will be at the merging point at a certain time $t_i^0 + (k+1)\Delta t$, i.e., $d_i(t_i^m) = L$ is hard to be strictly satisfied. Therefore, as long as the position satisfies $x_i(t_i^m) > L$ at a certain time $t_i^0 + (k+1)\Delta t$, then CAV $i$ is considered to have completed merging. After that, CAV $i$ switches to car-following mode. It is worth mentioning that when $\Delta t$ is small enough, the approximation errors of the HDV's state and the CAV's control input in the recursive interval are insignificant.

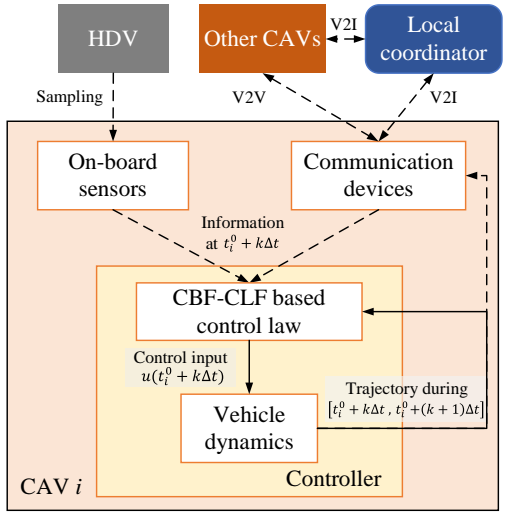

Figure 2. The recursive optimal control framework.

The formulated optimal control problem is usually solved by using optimal control methods, e.g., Pontryagin's maximum principle. However, as one or more constraints become active, there would be increasing computational complexity, and even leave the problem unsolved. Moreover, for a nonlinear vehicle dynamics system that is more accurate than a simple linear vehicle dynamics system, the classical optimal control methods are ineffective [17]. To tackle this problem, an optimal control method based on control barrier functions and control Lyapunov functions was introduced. By using CBFs and CLFs, we can map the safety constraints and objectives from the state $x_i(t)$ to control input $u_i(t)$, and form constrained quadratic programs (QPs) [13]. Then, the safety-critical optimal control problem for nonlinear systems can be solved in real time while yielding the optimal solution.

In the next two parts, we use CBFs and CLFs to reformulate the optimal control problem mentioned in Section II. Hard constraints which must be satisfied are constructed in the form of CBF. For control objectives, the minimum travel time can be converted by reaching the desired speed $v_d$ as soon as possible and maintaining it, hence can be achieved using a CLF. The objective of minimum energy consumption can be directly reflected in the optimization objective function.

### B. Control Barrier Functions Design

*Definition 1. Control barrier function (CBF)* [11]: Consider a set $\mathbb{C} := \{x \in \mathbb{R}^n : \mathcal{S}(x) \geq 0\}$ of a continuously differentiable function $\mathcal{S}: \mathbb{R}^n \to \mathbb{R}$, then a function $B(x): \mathbb{C} \to \mathbb{R}$ is a control barrier function if there exist class $\mathcal{K}$ function $\rho_1, \rho_2$ and $\gamma > 0$ such that

$$\frac{1}{\rho_1(\mathcal{S}(x))} \leq B(x) \leq \frac{1}{\rho_2(\mathcal{S}(x))} \tag{7a}$$

$$\sup_{u \in U}[L_f B(x) + L_g B(x)u] - \frac{\gamma}{B(x)} \geq 0 \tag{7b}$$

for all $x \in Int(\mathbb{C})$, where $L_f$ and $L_g$ denote the Lie derivatives for system (2). All control values satisfy

$$K_{cbf}(x) := \left\{ u \in U : L_f B(x) + L_g B(x)u - \frac{\gamma}{B(x)} \geq 0 \right\} \tag{8}$$

can render $\mathbb{C}$ safe. In this paper, we take $B(x) = 1/\mathcal{S}(x)$ satisfying (7) with $\rho_1(\mathcal{S}(x)) = \rho_2(\mathcal{S}(x)) = \mathcal{S}(x)$ and $\gamma = 1$.

For the maximum speed limitation of (6c), selecting $\mathcal{S}_{i,1}(x_i(t)) = v_{max} - v_i(t)$ yields the control barrier function $B_{i,1}(x_i(t)) = 1/\mathcal{S}_{i,1}(x_i(t))$, then the control law

$$L_f B_{i,1}(x_i(t)) + L_g B_{i,1}(x_i(t))u_i(t) \geq \frac{\gamma_{i,1}}{B_{i,1}(x_i(t))} \tag{9}$$
$$\forall t \in [t_i^0, t_i^m]$$

$$L_f B_{i,1}(x_i(t)) = \frac{-F_r(v_i(t))}{\delta M(v_{max} - v_i(t))^2}$$

$$L_g B_{i,1}(x_i(t)) = \frac{1}{\delta M(v_{max} - v_i(t))^2}$$

should be satisfied for any control input $u_i(t)$. Similarly, minimum speed limitation leads to $\mathcal{S}_{i,2}(x_i(t)) = v_i(t) - v_{min}$ and corresponding control law like (9).

The acceleration constraint is associated with control input through (2), thus derives time-varying constraints of $u_i(t)$

$$\delta M a_{min} + F_{r,i}(t) \leq u_i(t) \leq \delta M a_{max} + F_{r,i}(t) \tag{10}$$

Safe merging constraint (6c) only exists at the merging point, corresponding to time $t_i^m$, thus it is not a continuous time-varying constraint related to time. However, in order to construct a time-varying constraint form of CBF, we need to transform the safe merging constraint to a time-continuous form. In this paper, the time headway $\varphi$ is regarded as a time-varying function $\Phi(t)$. To make the safe merging constraint intervene in CAV control gradually and smoothly, we assume $\Phi(d_i(t))$ is linearly related to travel distance with $\Phi(d_i(t_i^0)) = 0$ and $\Phi(d_i(t_i^m)) = \varphi$. Let $l = 0$ for simplicity. Combined with $d_i(t_i^0) = 0$ and $d_i(t_i^m) = L$, then the time-varying time headway can be written in linear form as

$$\Phi(d_i(t)) = \frac{\varphi d_i(t)}{L} \tag{11}$$

therefore, (6c) is rewritten in a continuous form as

$$z_{i,i-1} \geq l + \Phi(d_i(t))v_{i-1}(t), \forall t \in [t_i^0, t_i^m] \tag{12}$$

The CBF of safe-merging can be constructed through $\mathcal{S}_{i,3}(x_i(t)) = z_{i,i-1} - l - \Phi(d_i(t))v_{i-1}(t)$. Since the control input constraint (10) may conflict with the constraint (12) especially when CAV decelerates, it is necessary to construct a control barrier function of safe merging constraint considering the influence of control input. Assuming vehicle $i$ decelerates with the minimum control input $u_i(t) = \delta M a_{min} + F_{r,i}(t)$, and neglecting resistance force $F_{r,i}(t)$ as it increases the braking force, we have $v_i(t + \tau) = v_i(t) + \delta M a_{min} \tau$ according to vehicle dynamics (2). In the case that the speed of vehicle $i$ decreases to $v_i(t + T) = v_{i-1}(t + T)$ during time interval $[t, t + T]$ and the minimum distance $z_{i,i-1\_min} = l + \Phi(d_i(t + T))v_i(t + T)$ is achieved, assuming $v_{i-1}(t)$ does not change during this time interval when calculating $T$ (as the recursive period is small), we have

$$T = \frac{v_i(t+T) - v_i(t)}{\delta M a_{min}} = \frac{v_{i-1}(t) - v_i(t)}{\delta M a_{min}} \tag{13}$$

Then the distance after time $T$ is

$$\begin{aligned}
z_{i,i-1}(t+T) &= z_{i,i-1}(t) + \int_0^T [v_{i-1}(t) - v_i(t+\tau)]d\tau \\
&= z_{i,i-1}(t) + \int_0^T [v_{i-1}(t) - v_i(t) - \tau\delta M a_{min}]d\tau \\
&= z_{i,i-1}(t) + \frac{1}{2}\frac{(v_{i-1}(t) - v_i(t))^2}{\delta M a_{min}}
\end{aligned} \tag{14}$$

The displacement of vehicle $i$ during time interval $[t, t+T]$ is

$$\begin{aligned}
x_i(t+T) - x_i(t) &= v_i(t)T + \frac{1}{2}\delta M a_{min} T^2 \\
&= -\frac{1}{2}\frac{(v_{i-1}(t) - v_i(t))^2}{\delta M a_{min}}
\end{aligned} \tag{15}$$

Substituting (14) and (15) into (12) gives

$$z_{i,i-1}(t) \geq l - \frac{1}{2}\frac{(v_{i-1}(t) - v_i(t))^2}{\delta M a_{min}} + \frac{\varphi(d_i(t) - \frac{1}{2}\frac{v_i^2(t) - v_{i-1}^2(t)}{\delta M a_{min}})v_i(t)}{L} \tag{16}$$

Hence, we choose

$$S_{i,4}(x_i(t)) = z_{i,i-1}(t) + \frac{1}{2}\frac{(v_{i-1}(t) - v_i(t))^2}{\delta M a_{min}} - l - \frac{\varphi(d_i(t) - \frac{1}{2}\frac{v_i^2(t) - v_{i-1}^2(t)}{\delta M a_{min}})v_i(t)}{L} \tag{17}$$

Similarly, for safe car-following, we choose

$$S_{i,5}(x_i(t)) = z_{i,ip}(t) - l - \varphi v_{ip}(t) \tag{18}$$

$$S_{i,6}(x_i(t)) = z_{i,ip}(t) + \frac{1}{2}\frac{(v_{ip}(t) - v_i(t))^2}{\delta M a_{min}} - \varphi v_i(t) - l \tag{19}$$

Then we can derive control laws for $S_{i,3}$, $S_{i,4}$, $S_{i,5}$, and $S_{i,6}$ in the CBF form.

### C. Control Lyapunov Function Design

Now we consider the objectives. The objective of minimizing travel time can be indirectly achieved by approaching the desired speed $v_d \in [v_{min}, v_{max}]$ as soon as possible using the control Lyapunov function.

*Definition 2. Control Lyapunov function (CLF)* [11]: Consider a continuously differentiable function $V: \mathbb{R}^n \to \mathbb{R}$, if there exist positive constants $\xi_1, \xi_2, \xi_3 > 0$ such that

$$\xi_1 \|x\|^2 \leq V(x) \leq \xi_2 \|x\|^2 \tag{20a}$$

$$\inf_{u \in U}[L_f V(x) + L_g V(x)u] + \xi_3 V(x) \leq 0 \tag{20b}$$

for $\forall x \in \mathbb{R}^n$, $\xi_3$ is the exponential convergence rate. For any Lipschitz continuous controller $u \in K_{clf}(x)$ with

$$K_{clf}(x) := \{u \in U: L_f V(x) + L_g V(x)u + \xi_3 V(x) \leq 0\} \tag{21}$$

the system (2) can be exponentially stabilized to its zero dynamics.

Define output $y_i(x_i) := v_i - v_d$ and choose Lyapunov function $V_i(y_i) = y_i^2$ satisfying (20) with $\xi_1 = \xi_2 = 1$ and $\xi_3 = \varepsilon > 0$. Then the corresponding control input law is

$$\underbrace{-\frac{2(v_i(t) - v_d)}{\delta M}F_r(v_i(t))}_{L_f V_i(y_i(t))} + \underbrace{\frac{2(v_i(t) - v_d)}{\delta M}u_i(t)}_{L_g V_i(y_i(t))} + \varepsilon(v_i(t) - v_d)^2 \leq \theta_i(t), \quad \forall t \in [t_i^0, t_i^m] \tag{22}$$

where $\theta_i(t) > 0$ is a slack variable to make (22) a soft constraint. By optimizing $\theta_i(t)$, the travel efficiency can be indirectly improved. Combining with (6a), we have a new objective function

$$J_i' = v_i u_i + c' u_i^2 + w_i \theta_i \tag{23}$$

### D. Quadratic Programming Problem Formulation

As discussed in Section III.A, Problem I is recursively solved to tackle the disturbances. Hence, we can transform Problem I with the new objective function (23) into a discrete quadratic programming (QP) Problem.

*Problem II. Discrete quadratic programming problem*

$$q_i^* = \arg\min_{q_i} \frac{1}{2}q_i^T L_i q_i + H_i^T q_i \tag{24}$$

$$q_i = \begin{bmatrix} u_i \\ \theta_i \end{bmatrix} \quad L_i = \begin{bmatrix} 2cR^2/\mathcal{T}^2 & 0 \\ 0 & 2\omega_i \end{bmatrix} \quad H_i = \begin{bmatrix} v_i \\ 0 \end{bmatrix}$$

Subject to

$$L_f B_{i,k}(x) + L_g B_{i,k}(x)u_i(t) - \frac{\gamma}{B_{i,k}(x)} \geq 0 \quad \text{Constraints } (k = 1 \dots 6)$$

$$L_f V_i(x) + L_g V_i(x)u_i(t) + \xi_3 V_i(x) \leq \theta_i \quad \text{Objective}$$

## IV. SIMULATION AND RESULTS

### A. Three-vehicle scenario simulation

Consider an on-ramp merging scenario among three vehicles. The first vehicle is an HDV on the main road, and the other two vehicles are CAVs on the on-ramp and main road, respectively. According to the principle of first in first out (FIFO), the initial conditions of vehicles are given in Table II.

TABLE II.   INITIAL CONDITIONS

| Road | Vehicle denotation | Initial position (m) | Initial speed (m/s) |
|---|---|---|---|
| Main road | HDV | Earlier than CAV 1 | / |
| On-ramp | CAV 1 | 0 | $v_2(0) = 10$ |
| Main road | CAV 2 | 0.9 | $v_3(0.9) = 19$ |

For the first HDV, we assume that its trajectory follows the law of trigonometric function and can be expressed as (25).

$$d_1(t) = 97 + 20t + \frac{8}{\pi}\cos\left(\frac{\pi}{10}t\right) - \frac{8}{\pi} \tag{25a}$$

$$v_1(t) = 20 - \frac{4}{5}\sin\left(\frac{\pi}{10}t\right) \tag{25b}$$

When $t = 0$, the initial position $d_1(t_0)$ is 97m and initial velocity $v_1(t_0)$ is 20m/s.

The main parameters of CAVs are listed in Table I. For (22), take the exponential convergence rate $\varepsilon = 10$, the proportion of this objective in the objective function (23) is $\omega_i = 10$. CAVs need to repeatedly collect the HDV's information and recursively plan their own trajectories. In this paper, the sampling time is $\Delta t = 0.1s$. The desired speed of CAVs is 30m/s and the length of the coordination zone is 400m. In the mixed traffic, when a vehicle's position exceeds 400m, then the vehicle is judged to merge successfully.

The simulation results are shown in Figure 3. Generally, the on-ramp vehicle CAV1 merges at 20.9s successfully after the merging control process imposed on CAV1 and CAV2. The inter-vehicle distance requirements (6d), (6e), acceleration limitations (6b), and the corresponding traction force constraints (11) are all satisfied during the whole process. More specifically, at the beginning, the inter-vehicle distances are improper, i.e., the distance between the leading HDV and CAV1 is too big, and CAV1 is very close to CAV2. Hence, CAV1 accelerates with the maximum acceleration to catch up HDV as soon as possible, while CAV2 first decelerates with minimum deceleration for 0.2 seconds to keep an inter-vehicle distance from CAV1, then decelerate more gently and accelerates for a while. After 4s, both CAV1 and CAV2 hold steady speeds to shorten distances from the corresponding preceding vehicles gradually. After 8s, since the distance of HDV-CAV1 approaches the distance requirement (6c) as shown in Figure 3(d), CAV1 decelerates to maintain the distance, which also makes CAV2 decelerate. At around 12s, it's obvious that both inter-vehicle distances satisfied merging requirements and the relative speeds are nearly zero, thus it can be a proper opportunity to complete merging. However, since CAVs does not reach the merging position 400m, they still need to follow the on-ramp merging control demands. As the time-varying time headway function gets bigger, the required inter-vehicle distances get bigger, resulting in CAVs decelerating to generate required merging distance. Finally, CAV1 merges with proper merging distance.

Figure 3(f) shows the battery energy consumption during the merging control process, which changes with the pace of speeds. Note that, after 8s, the consumption decrease is caused by regenerative braking. In terms of computational performance, we use MATLAB to conduct all simulations on a desktop computer with an Intel ® CORE$^{TM}$ i7-9700K CPU @ 3.60GHz, the computation time of each step is less than 0.004s, which proves that the proposed method can be applied in real time.

*B. Multi-HDV scenario simulation*

We also conduct simulations to verify the proposed control method in multi-HDV scenarios. Three different scenarios are compared: (i) full-HDV scenario, (ii) three on-ramp CAVs are controlled by the proposed control method in mixed traffic, and (iii) three on-ramp and the corresponding following main-lane CAVs are controlled by the proposed control method in mixed traffic. We put 30 vehicles into each simulation scenario. All main-lane or on-ramp vehicles arrive at the initial position with speeds of 20m/s or 15m/s. The leading HDV drives with a given trajectory. All the following HDVs are modelled by widely-used Intelligent Driver Model (IDM) [16]. For CAVs, they switch to the car-following operation after merging control process.

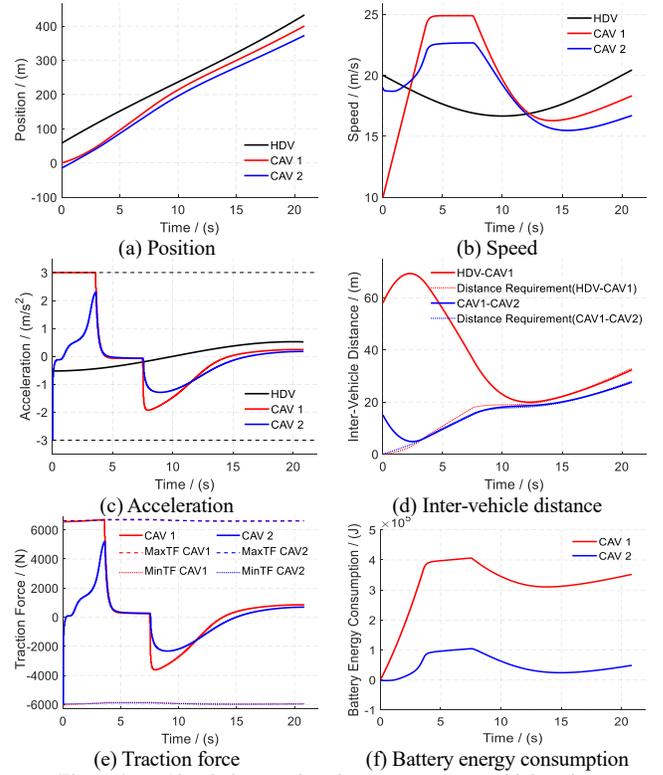

Figure 3. Simulation results when one on-ramp vehicle exists.

Speed trajectories are shown in Figure 4. It is shown that although the CAVs may have more violent acceleration and deceleration operations than HDVs, their existence actually limits the whole speed variation ranges of all vehicles. As Figure 4(a) shows, in full-HDV traffic, the speed reduction caused by the leading HDV spread to the upstream traffic, which can cause serious traffic jam in the merging bottleneck. However, if there are CAVs controlled by the proposed merging control method exist in the merging area, the whole speed reduction is alleviated and hence the traffic efficiency can be improved.

Table III presents average travel time and battery energy consumption of all 30 vehicles when each of them pass the merging position 400m. The results reveal that the proposed method applied on CAVs effectively shortens the overall merging duration, and the saving ratio increases as the number of CAVs increases because the speed reduction is mitigated. In terms of battery energy consumption, the proposed method also performs better than the full-HDV scenario. Note that the consumption of scenario (iii) is more than that of scenario (ii) to keep higher vehicle speeds.

V. CONCLUSIONS

In this paper, a decentralized optimal control method is introduced to solve the on-ramp merging problem of CAVs in the mixed traffic. The objectives of improving travel efficiency and energy efficiency are constructed into CLFs, and the control and safety-critical constraints are constructed into CBFs. Based on a recursive optimal control framework,

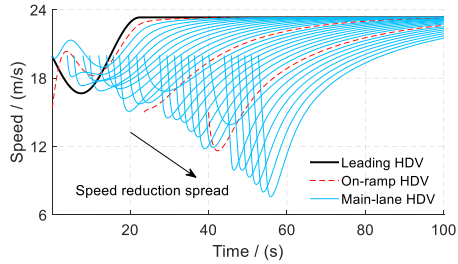

(a) Scenario (i): Full-HDV scenario.

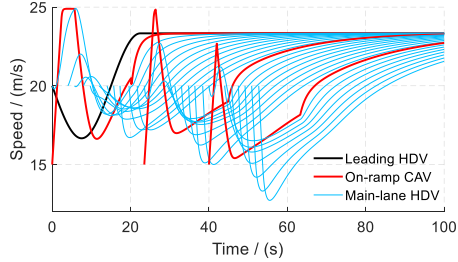

(b) Scenario (ii): Three on-ramp CAVs are controlled by the proposed control method in mixed traffic.

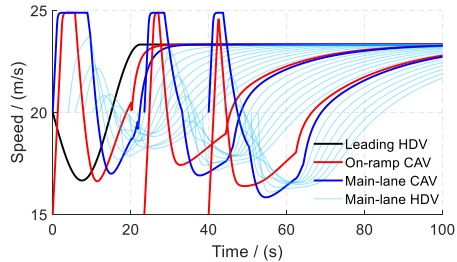

(c) Scenario (iii): Three on-ramp and the corresponding following main-lane CAVs are controlled by the proposed control method in mixed traffic.

Figure 4. Simulation results when multiple HDVs exist.

TABLE III. SIMULATION RESULTS OF MULTI-HDV SCENARIOS

| Scenario | (i) | (ii) | (iii) |
|---|---|---|---|
| *Average travel time (s)* | 23.3689 | 22.3896 | 21.5931 |
| *Saving ratio* | / | 4.19% | 7.60% |
| *Average battery energy consumption (x $10^6$ J)* | 1.3626 | 0.7320 | 1.0438 |
| *Saving ratio* | / | 46.28% | 23.40% |

CAVs can plan trajectories while dealing with uncertain disturbances of HDVs in real time. Numerous simulation results show that the proposed method have the potential to improve vehicle travel efficiency and energy-saving effects while ensuring driving safety, control stability and robustness.